\documentstyle[prc,aps]{revtex}

\begin{document}
\title{Asymptotic Normalization Coefficients for $^{13}C + p \rightarrow {}^{14}N$}
\author{L. Trache, A. Azhari, H.~L. Clark, C.~A. Gagliardi, Y.-W. Lui, A.~M.
Mukhamedzhanov, and R.~E. Tribble}
\address{Cyclotron Institute, Texas A\&M University, College\\
Station, \\
TX 88743-3366, U.S.A.}
\author{F. Carstoiu}
\address{Institute of Physics and Nuclear Engineering, Bucharest,\\
Romania}
\date{\today}
\maketitle

\begin{abstract}
The $^{13}C(^{14}N,^{13}C)^{14}N$ proton exchange reaction has been measured
at an incident energy of 162 MeV. Angular distributions were obtained for
proton transfer to the ground and low lying excited states in $^{14}N$.
Elastic scattering of $^{14}N$ on $^{13}C$ also was measured out to the
rainbow angle region in order to find reliable optical model potentials.
Asymptotic normalization coefficients for the system $^{13}C + p \rightarrow
{}^{14}N$ have been found for the ground state and the excited states at
2.313, 3.948, 5.106 and 5.834 MeV in $^{14}N$. These asymptotic
normalization coefficients will be used in a determination of the S-factor
for $^7Be(p,\gamma)^8B$ at solar energies from a measurement of the proton
transfer reaction $^{14}N(^{7}Be,^{8}B)^{13}C$.
\end{abstract}

\section{Introduction}

The asymptotic normalization coefficient $C$ for the system $A + p
\leftrightarrow B$ specifies the amplitude of the single-proton tail of the
wave function for nucleus $B$ when the core $A$ and the proton are separated
by a distance large compared to the strong interaction radius. In previous
reports \cite{ANC0,Xu1}, we have shown that knowledge of asymptotic
normalization coefficients (ANC's) can be used to calculate the direct
capture rates for $(p,\gamma)$ or $(\alpha,\gamma)$ reactions of
astrophysical interest when the captured $p$ or $\alpha$ is relatively
loosely bound in the final nucleus. The required ANC's can often be measured
in peripheral transfer reactions. We are using the ANC technique to
determine the astrophysical S-factor $S_{17}(0)$ for the proton radiative
capture reaction $^{7}Be(p,\gamma )^{8}B$ at solar energies, using the
transfer reactions $^{10}B(^{7}Be,^{8}B)^{9}Be$ and $%
^{14}N(^{7}Be,^{8}B)^{13}C$. In order to extract the ANC for $^{7}Be +
p\rightarrow {}^{8}B$ from these measurements, we must know the ANC's for
the $^{9}Be + p \rightarrow {}^{10}B$ and $^{13}C + p \rightarrow {}^{14}N$
systems. We report below a study of $^{14}N$ + $^{13}C$ elastic scattering
and the proton exchange reaction $^{13}C(^{14}N,^{13}C)^{14}N$ at 162 MeV,
from which we find the ANC's corresponding to $^{13}C + p \rightarrow
{}^{14}N$. The experiment is similar to our measurement of the $^{9}Be + p
\rightarrow {}^{10}B$ ANC's reported earlier \cite{B10ANC}.

Below we present details of the experiment. This is followed by a discussion
of the optical model parameters extracted from the elastic scattering data
and then the results for the ANC's found from the proton exchange reaction.

\section{The Experiment}

The experiment was performed using a $^{14}N$ beam from the Texas A\&M
University K500 superconducting cyclotron and the Multipole Dipole Multipole
magnetic spectrometer \cite{MDM}. A 300 $\mu$g/cm$^{2}$ self-supporting
target of 99\% enriched $^{13}C$ was bombarded with a well collimated 162
MeV $^{14}N^{+3}$ beam. The angular spread of the beam on target was less
than $\Delta \theta =0.1^{\circ}$ after passing through the Beam Analysis
System \cite{BAS}. Both elastic scattering and the proton transfer reaction
were measured during the same run. The elastic scattering data were used to
assess the possible effects of interference between the elastic scattering
and exchange processes and to extract optical model parameters for use in
the DWBA calculations of the proton exchange reaction. The elastic
scattering results were also used in the normalization of the cross sections
for the transfer reaction. The experimental setup was identical to that used
in the $^{10}B+{}^{9}Be$ experiment and was described in detail in \cite
{B10ANC}. For the present experiment, the spectrometer's entrance aperture
was set at $\Delta\theta =4^{\circ}$ (horizontal) and $\Delta \varphi
=1^{\circ}$ (vertical). The modified Oxford detector \cite{Oxfdet} was used
in the focal plane. The detector consists of a 50 cm long gas ionization
chamber to measure the specific energy loss of particles in the gas and
their focal plane position at four resistive wires, separated by 16 cm along
the particles' trajectories, followed by an NE102A plastic scintillator to
measure the residual energy. The entrance and exit windows of the detector
were made of 1.8 and 7.2 mg/cm$^{2}$ thick Kapton foils, respectively. The
ionization chamber was filled with purified isobutane at a pressure of 30
Torr.

Elastic scattering data were obtained over the laboratory angular range $%
\theta_{lab}=2^{\circ} - 34^{\circ}$, corresponding to the center-of-mass
range $\theta _{cm}=4^{\circ} - 70^{\circ}$, by detecting $^{14}N^{+7}$ in
the focal plane of the spectrometer. The proton exchange reaction was
measured by retuning the magnetic fields of the spectrometer for the
rigidity of the outgoing $^{13}C^{+6}$ in the forward angle range $\theta
_{lab}=-3^{\circ}$ to $+18^{\circ}$. This is kinematically equivalent to
measuring elastic or inelastic scattering at the complementary backward
angles. Particle identification was accomplished by using the energy loss
measured in the ionization chamber and the residual energy as determined by
the light output from the plastic scintillator. The focal plane position and
the scattering angle at the target were reconstructed using the position
measurements from any two of the four wires in the detector, coupled with
RAYTRACE \cite{raytrace} calculations. Typically we used the position at the
first wire in the detector and that at the wire closest to the focal plane.
The spectrometer angular acceptance range of 4$^{\circ}$ was divided into 8
bins of 0.5$^{\circ}$ each during the data analysis. As a check on the
reconstruction, we calibrated the target scattering angle determination
using an angle mask with five slits $\Delta \theta =0.1^{\circ}$ wide,
uniformly distributed across the 4$^{\circ}$ opening. These measurements
also indicated that the total angular resolution for the experiment was $%
\Delta \theta _{lab}=0.2^{\circ}$. The low-lying excited states in both $%
^{13}C$ and $^{14}N$ are well known. Thus, the focal plane energy
calibration was straightforward. Typically the spectrometer was moved in 3$%
^{\circ}$ steps, allowing for an angle overlap between measurements to check
for consistency in the results. Due to the high purity of the target,
elastic scattering data were obtained down to $\theta _{lab}=2.5^{\circ}$
without contamination from heavier elements in the target. By combining
measurements of the target thickness with the normalization to elastic
scattering at very forward angles, the absolute cross sections for the
proton transfer reactions have been determined with an uncertainty of 7\%. A
spectrum for the proton transfer reaction taken at $\theta_{lab}=8^{\circ}$
is shown in Fig.~\ref{fig1}. In addition to transfer between the ground
states of $^{14}N$ and $^{13}C$ (elastic proton exchange), we see
transitions populating the first (2.313 MeV, $J^{\pi}$=0$^{+}$, $T$=1),
second (3.948 MeV, $J^{\pi}$=1$^{+}$, $T$=0), fourth (5.106 MeV, $J^{\pi}$=2$%
^{-}$, $T$=0) and sixth (5.834 MeV, $J^{\pi}$=3$^{-}$, $T$=0) excited states
of $^{14}N$ and the first excited state of $^{13}C$ (3.089 MeV, $J^{\pi}$=1/2%
$^{+}$), where excitation energies, spins and parities have been taken from 
\cite{ajselove}.

\section{Optical Model Potentials}

The measured elastic scattering angular distribution is shown in Fig.~\ref
{fig2}. Data at forward angles are from normal kinematics elastic
scattering, while the data at back angles have been taken from the $%
^{13}C(^{14}N,^{13}C)^{14}N$ reaction at forward angles, populating the
ground states of both $^{13}C$ and $^{14}N$. While the forward angle data
involving proton exchange are kinematically equivalent to elastic scattering
at back angles, it is clear from the figure that potential scattering and
the proton transfer mechanism completely dominate at forward and backward
angles, respectively. We thus treat the data in the two angular ranges
independently and do not consider any interference between the amplitudes of
the two processes.

The forward angle data have been fit using the code OPTIMIX \cite{Optimix}
in a standard optical model analysis using Woods-Saxon volume form-factors
for the potential: 
\begin{equation}
U(r)=-(Vf_V(r)+iWf_W(r)),  \label{potential}
\end{equation}
with the usual notation where 
\begin{equation}
f_{x}(r)=\left[ 1+\exp \frac {r-r_{x}(A_{1}^{1/3}+A_{2}^{1/3})}{a_{x}}
\right] ^{-1}.  \label{formf}
\end{equation}
$V$ and $W$ are the depths of the real and imaginary potentials, $A_{1}$ and 
$A_{2}$ are the nuclear masses, $r_{x}$ and $a_{x}$ are the reduced radii
and diffuseness of the potentials, and $x$ can be either $V$ or $W$ for the
real and imaginary parts of the potentials, respectively. Only the central
potential terms have been included since vector and higher rank tensor
spin-orbit couplings have negligible impact on the cross sections.

Five distinct families of potentials were found in the chi square analysis
of the data. Their parameters are presented in Table I, and the fits are
compared with the forward angle data in Fig.~\ref{ratruth}. Included in the
table are the volume integrals per pair of interacting nucleons for the real
and imaginary parts of the potentials (J$_V$ and J$_W$), their rms radii (R$%
_V$ and R$_W$), and the total reaction cross section calculated in the
Glauber model. We note that the volume integrals increase regularly from one
family to the next, indicating that no family was missed during the
automatic search for the minima. The five potential sets reproduce the total
reaction cross section $\sigma_{R}$ = 1463(100) mb measured by DiGregorio 
{\it et al}. at 161.3 MeV \cite{totalcsec}. All of the potentials give
reasonable $\chi^2$, but potential P1 listed in the table gives the smallest
value and is the only one that fits the data at largest angles. This
potential also has a real volume integral per pair of interacting nucleons
close to that we found (206 MeV$\cdot$fm$^3$) for the preferred potential in
our previous study of $^{10}B$ + $^9Be$ elastic scattering at similar
velocities \cite{B10ANC}. Hence, we have adopted potential P1 for the DWBA
calculations of the proton transfer process, while the others are used to
estimate the uncertainty due to the choice of optical model parameters.
Further details concerning the potential model analysis will be discussed in
a future publication.

\section{Asymptotic Normalization Coefficients}

For a peripheral transfer reaction, ANC's are extracted from the measured
angular distribution by comparison to a DWBA calculation. Consider the
proton transfer reaction $a + A \to c + B$, where $a=c + p$ and $B= A + p$.
The experimental cross section is related to the DWBA according to 
\begin{equation}
\frac{d\sigma}{d\Omega}= \sum_{l_B j_B l_a j_a} (C^{B}_{A p l_B j_B})^{2}
(C^{a}_{c p l_a j_a})^{2} R_{l_B j_B l_a j_a} ,  \label{dwcs2}
\end{equation}
where 
\begin{equation}
R_{l_B j_B l_a j_a}= \frac{{\tilde \sigma}_{l_B j_B l_a j_a}^{DW}} {b^{2}_{A
p l_B j_B} b^{2}_{c p l_a j_a}} .  \label{r1}
\end{equation}
${\tilde \sigma}$ is the calculated DWBA cross section and the $b$'s are the
asymptotic normalization constants for the single particle bound state
orbitals used in the DWBA. The sum in Eq. (\ref{dwcs2}) is taken over the
allowed orbital and total angular momentum couplings, and the $C$'s are the
ANC's for $a \to c + p$ and $A + p \to B$. For peripheral proton transfer,
the above normalization of the DWBA cross section by the ANC's for the
single particle orbitals makes the extraction of the ANC for $A + p \to B$
essentially independent of the parameters used in the single particle
potential wells, in marked contrast to the more typical parametrization of
the DWBA cross section in terms of spectroscopic factors. See \cite{B10ANC}
for additional details.

The angular distribution for the proton exchange reaction involving both the
target and projectile ground states -- elastic proton transfer -- is shown
in Fig.~\ref{ang1}. DWBA calculations for the proton transfer were carried
out with the finite-range DWBA code PTOLEMY \cite{Ptolemy}, using the full
transition operator. Distorted waves were calculated using optical model
potential P1 in Table I, and a standard Woods-Saxon well was used to bind
the transferred proton to the remaining nuclear core. As was noted above,
the spectroscopic factor associated with elastic transfer differs from the
ANC by the normalization of the single particle wave function ANC's
calculated in the same Woods-Saxon well. If a reaction is peripheral, this
makes the extracted ANC quite stable over a broad range of single particle
well parameters. In Fig.~\ref{spect}, we compare the ground state
spectroscopic factor $S_{p_{1/2}}$ and ANC $C_{p_{1/2}}^2$ extracted for
parameters of the single particle potential ranging from $r_{0}=1.0-1.3$ fm
and $a=0.5-0.7$ fm, as functions of the value of the corresponding single
particle ANC, $b_{p_{1/2}}$. It is clear from the figure that the
spectroscopic factor depends strongly on the choice of the single particle
potential parameters, while the ANC varies by less than 7\% over the full
range. If the choice of single particle well parameters is constrained to be
within reasonable agreement with the measured rms charge radius \cite{radii}%
, the variation of the ground state ANC $C_{p_{1/2}}^2$ is less than 3\%
whereas the spectroscopic factor varies by over 25\%. A similar picture
arises for $S_{p_{3/2}}$ and $C_{p_{3/2}}^2$, despite a substantially
smaller contribution of the $1p_{3/2}$ orbital to the proton transfer cross
section, and we take this as a confirmation of our fits for $C_j^2$. Another
indication of the peripheral character of the reaction is the localization
of the transfer strength with partial waves. For the elastic transfer, the
DWBA transition matrix element is peaked around $l$ values of 32, which
corresponds semiclassically to $r$ = 6.4 fm, and has a FWHM of about 10,
making this reaction even more strongly focused on the surface than the $%
^{9}Be(^{10}B,^{9}Be)^{10}B$ elastic transfer reported in \cite{B10ANC}.

From Eq. (\ref{dwcs2}), the elastic proton transfer cross section is
proportional to $C^4$ since the entrance and exit channels are identical.
For the elastic transfer, we assumed a mixed configuration for the ground
state of $^{14}N$ ($J^{\pi}$=1$^{+}$, $T$=0) in which the last proton in
either the $1p_{1/2}$ or $1p_{3/2}$ orbital is coupled to the 1/2$^{-}$
ground state of $^{13}C$ . Only the $1p_{1/2} \rightarrow 1p_{1/2}$ and $%
1p_{1/2} \leftrightarrow 1p_{3/2}$ contributions were considered since the
admixture of the $1p_{3/2}$ orbital is small and the calculated angular
distribution for $1p_{3/2} \rightarrow 1p_{3/2}$ is virtually
indistinguishable from that for $1p_{1/2} \rightarrow 1p_{1/2}$. Note that
the $1p_{1/2} \rightarrow 1p_{3/2}$ and $1p_{3/2} \rightarrow 1p_{1/2}$
contributions are identical due to time reversal invariance. Core
excitations were not included since they should give a negligible
contribution to the direct proton exchange. The DWBA calculation is compared
to the data in Fig.~\ref{ang1}. The solid line was found by combining
contributions from the $1p_{1/2}$ and $1p_{3/2}$ components, weighted by the
extracted $C^2$ for each $j$ transfer. The extracted elastic transfer ANC's
are given in Table II. The uncertainties in the extraction of the dominant $%
C_{p_{1/2}}^2$ term include the normalization of the cross section (3.5\%),
the choice of optical model parameters (3\%), the stability of the fits as a
function of the angular range considered (4\%), and the choice of
Woods-Saxon well parameters (1.5\%). In particular, we found that the
calculated DWBA transfer cross sections varied by only $\approx$2\% when
going from one family of optical model parameters to the next, and therefore 
$C^2$ changed by only half that. This insensitivity of the ANC to the choice
of optical model potential provides further support for the peripheral
nature of the $^{13}C(^{14}N,^{13}C)^{14}N$ reaction at this energy because
the elastic scattering was fitted in the angular range where it is
essentially diffractive in nature and the potential at the surface is well
determined.

In addition to the ground state, four of the excited states shown in Fig.~%
\ref{fig1} were populated with sufficient statistics to extract ANC's. The
one exception is the $^{13}C$ excited state at 3.089 MeV. We assume this
state was populated by removing a $2s_{1/2}$ proton from the small $%
2s_{1/2}^2$ component of the $^{14}N$ ground state. At small angles where
the $^{13}C$ excited state was clearly visible, the observed angular
distribution is consistent with a $2-3$\% admixture of this configuration in
the $^{14}N$ ground state.

The angular distributions for transitions to the $^{14}N$ excited states at
2.313, 3.948, 5.106 and 5.834 MeV are shown in Fig.~\ref{ang2}, together
with their calculated DWBA fits. In each case, the calculation was carried
out by considering the transition from the $^{14}N$ ground state to the
final proton configuration shown in Table II and yielded the ANC specified.
For the first excited state in $^{14}N$, the DWBA fit shown in Fig.~\ref
{ang2} includes $1p_{1/2} \rightarrow 1p_{1/2}$ and $1p_{3/2} \rightarrow
1p_{1/2}$ proton transfer terms, weighted by the $C_{p_{1/2}}^2$ and $%
C_{p_{3/2}}^2$ ANC's found above for the $^{14}N$ ground state,
respectively. A separate fit which allowed these two terms to vary
independently gave a result for $C_{p_{3/2}}^2/C_{p_{1/2}}^2$ for the ground
state that was consistent with the value found above, but with reduced
precision. For the second excited state, we considered contributions from $%
1p_{1/2} \rightarrow 1p_{1/2}$, $1p_{1/2} \rightarrow 1p_{3/2}$, and $%
1p_{3/2} \rightarrow 1p_{1/2}$ proton transfers. The latter two gave similar
calculated angular distributions and were combined. We found the $1p_{1/2}
\leftrightarrow 1p_{3/2}$ contribution to be very much smaller than the $%
1p_{1/2} \rightarrow 1p_{1/2}$ term given in Table II. The $^{14}N$ third
and fifth excited states, which form the $(1p_{1/2}\cdot$$%
2s_{1/2})_{0^-,1^-} $ doublet, were only weakly populated due to the angular
momentum mismatch and could not be resolved from the fourth and sixth
excited states, respectively. The $^{14}N$ fourth and sixth excited states
are members of the $(1p_{1/2}\cdot$$1d_{5/2})_{2^-,3^-}$ doublet, and their
characteristic oscillations are well described by the calculated $1p_{1/2}
\rightarrow 1d_{5/2}$ angular distribution. However, a 0.7$^{\circ}$ shift
is observed between the measured and calculated oscillations. Attempts to
include $1p_{3/2} \rightarrow 1d_{5/2}$ or $1p_{1/2} \rightarrow 2s_{1/2}$
terms, the latter to account for the weak unresolved states, did not improve
the fits. A similar situation, but with a shift of 2$^{\circ}$, was seen in
a previous $^{13}C(^7Li,^6He)^{14}N$ proton transfer experiment \cite{Li7He6}%
. As was noted above for the elastic transfer, the ANC's extracted for
transfer to the excited states depend only weakly on the assumed bound state
parameters or the choice of optical model potential. The uncertainties
quoted in Table II for the excited state ANC's are determined primarily by
the uncertainty in the normalization of the cross section (3.5\%) and the
added uncertainties due to the choice of optical model potential parameters
(3\%) and the quality and stability of the fits (4\% or larger). It is worth
noting that the normalization and optical potential uncertainties are
correlated for all of the ANC's in Table II.

\section{Conclusions}

We have measured the elastic scattering $^{13}C(^{14}N,^{14}N)^{13}C$ and
the elastic and inelastic proton exchange reaction $%
^{13}C(^{14}N,^{13}C)^{14}N$ leading to the ground state and four excited
states in $^{14}N$. The measurements of the proton transfer reaction have
been used to extract the ANC's describing the tail of the wave function of
the outer proton in $^{14}N$ in the field of the $^{13}C$ core. The ANC's
found here will be used to extract the ANC for $^{7}Be + p\rightarrow
{}^{8}B $ from the proton transfer reaction $^{14}N(^{7}Be,^{8}B)^{13}C$.

\acknowledgments

This work was supported in part by the U.S. Department of Energy under Grant
number DE-FG05-93ER40773 and by the Robert A. Welch Foundation.

\begin{table}[tbp]
\caption{ The parameters of the Woods-Saxon optical model potentials
extracted from the analysis of the elastic scattering data for $^{14}N$ (162
MeV) + $^{13}C$. $r_{C}$ = 1 fm for all potentials.}
\label{Tab1}
\begin{tabular}{|c|c|c|c|c|c|c|c|c|c|c|c|c|}
Pot. & $V$ & $W$ & $r_{V}$ & $r_{W}$ & $a_{V}$ & $a_{W}$ & $\chi^{2}$ & $%
\sigma_{R}$ & J$_{V}$ & R$_{V}$ & J$_{W}$ & R$_{W}$ \\ 
& [MeV] & [MeV] & [fm] & [fm] & [fm] & [fm] &  & [mb] & [MeV$\cdot$fm$^{3}$]
& [fm] & [MeV$\cdot$fm$^{3}$] & [fm] \\ \hline
P1 & 79.22 & 30.27 & 0.96 & 1.05 & 0.76 & 0.72 & 17.4 & 1542 & 221 & 4.52 & 
104 & 4.69 \\ 
P2 & 134.76 & 35.23 & 0.88 & 1.05 & 0.75 & 0.67 & 18.3 & 1525 & 299 & 4.28 & 
120 & 4.61 \\ 
P3 & 176.03 & 35.84 & 0.86 & 1.07 & 0.72 & 0.65 & 23.3 & 1527 & 361 & 4.15 & 
125 & 4.62 \\ 
P4 & 241.36 & 37.45 & 0.82 & 1.06 & 0.71 & 0.66 & 27.5 & 1533 & 438 & 4.00 & 
129 & 4.61 \\ 
P5 & 306.44 & 39.14 & 0.81 & 1.05 & 0.68 & 0.68 & 36.1 & 1552 & 522 & 3.90 & 
132 & 4.61
\end{tabular}
\end{table}

\begin{table}[tbp]
\caption{The Asymptotic Normalization Coefficients for the ${}^{13}C + p \to
{}^{14}N$ system, populating the ground and four excited states in $^{14}N$.
The calculations were done for the proton transferred from the ground state
of the $^{14}N$ projectile to the ``final proton configuration" in the
specified $^{14}N$ states.}
\label{Tab2}
\begin{tabular}{|c|c|c|c|}
State in & $J^{\pi}$,$T$ & Final proton & $(C_{lj})^2$ \\ 
$^{14}N$ &  & configuration & [fm$^{-1}$] \\ \hline
g.s. & 1$^{+}$,0 & $1p_{1/2}$ & 18.6(12) \\ 
&  & $1p_{3/2}$ & 0.93(14) \\ 
2.313 & 0$^{+}$,1 & $1p_{1/2}$ & 8.9(9) \\ 
3.948 & 1$^{+}$,0 & $1p_{1/2}$ & 2.8(3) \\ 
5.106 & 2$^{-}$,0 & $1d_{5/2}$ & 0.40(3) \\ 
5.834 & 3$^{-}$,0 & $1d_{5/2}$ & 0.19(2)
\end{tabular}
\end{table}

\begin{figure}[tbp]
\caption{Spectrum of the proton exchange reaction $%
^{13}C(^{14}N,^{13}C)^{14}N$ measured at $\theta_{lab}=8^{\circ }$. }
\label{fig1}
\end{figure}

\begin{figure}[tbp]
\caption{ The angular distribution for elastic scattering of $^{14}N$ on $%
^{13}C$. The data in the forward hemisphere were obtained by measuring the
elastically scattered $^{14}N^{+7}$, while those in the backward hemisphere
were obtained by measuring the transfer reaction product $^{13}C^{+6}$ at
the complementary forward angles. The dashed curve shows the Rutherford
scattering cross section, and the solid curve shows the cross section
calculation with potential P1 of Table I.}
\label{fig2}
\end{figure}

\begin{figure}[tbp]
\caption{The angular distribution for elastic scattering of 162 MeV $^{14}N$
on $^{13}C$ at forward angles. The curves are fits to the forward angle data
using the optical model potentials P1 (solid), P2 (dashed), P3 (dotted), P4
(dash-dotted), and P5 (solid) of Table I.}
\label{ratruth}
\end{figure}

\begin{figure}[tbp]
\caption{The angular distribution measured for the elastic proton exchange
reaction $^{13}C(^{14}N,^{13}C)^{14}N$. The curves show the DWBA fit over
the angular range $\theta_{cm}=0-12^{\circ}$ (full line), with $%
1p_{1/2}\rightarrow 1p_{1/2}$ (dashed line) and $1p_{1/2} \leftrightarrow
1p_{3/2}$ (dotted line) components. }
\label{ang1}
\end{figure}

\begin{figure}[tbp]
\caption{ The comparison between the spectroscopic factor $S_{p_{1/2}}$
(dots) and the ANC $C_{p_{1/2}}^2$ (squares) extracted for the ground state
of $^{14}N$ as a function of the single particle ANC, $b_{p_{1/2}}$, used to
normalize the DWBA calculations. Note that $C_{p_{1/2}}^2$ has been
multiplied by 0.1.}
\label{spect}
\end{figure}

\begin{figure}[tbp]
\caption{ The angular distributions for inelastic proton transfer to the $%
^{14}N$ excited states at 2.313, 3.948, 5.106 and 5.834 MeV, multiplied by
factors of $10^3$, $10^2$, 10 and 1, respectively. The curves show the
corresponding DWBA fits, as described in the text.}
\label{ang2}
\end{figure}

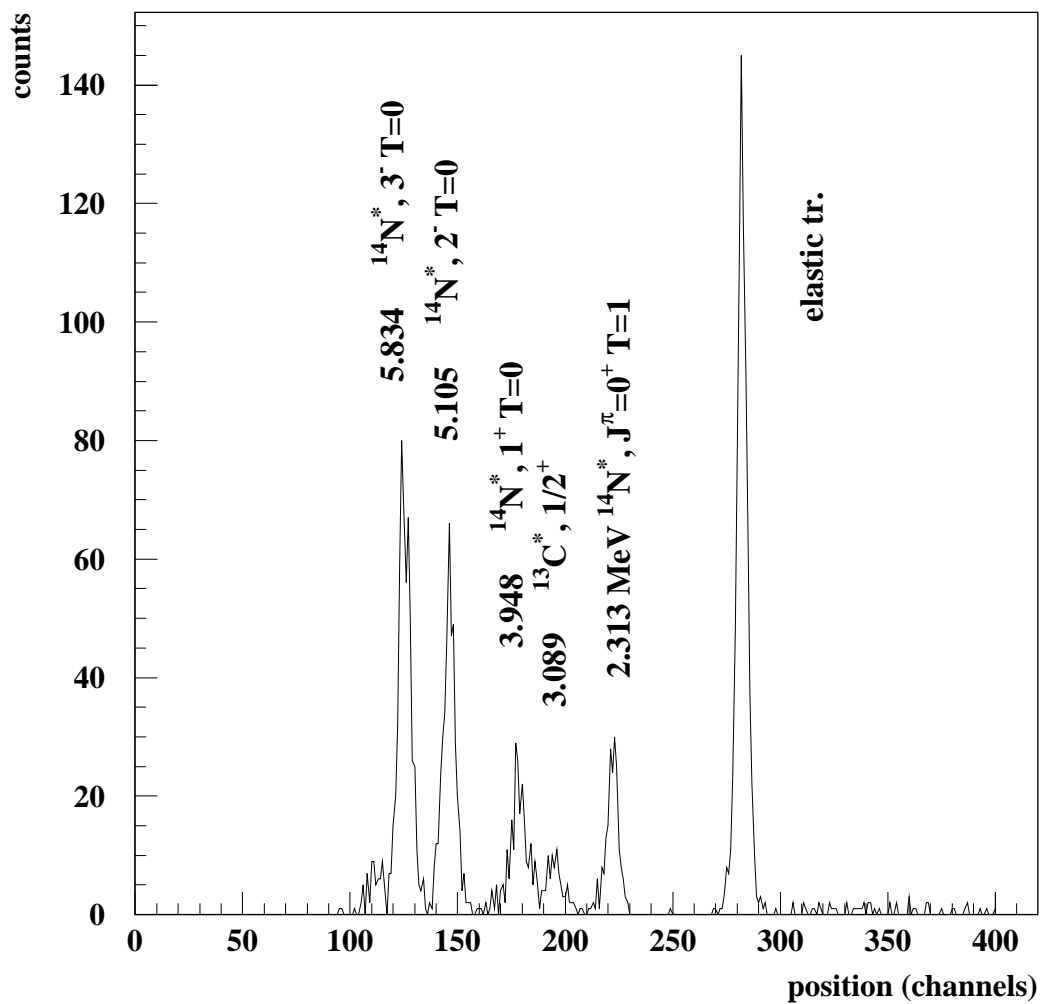
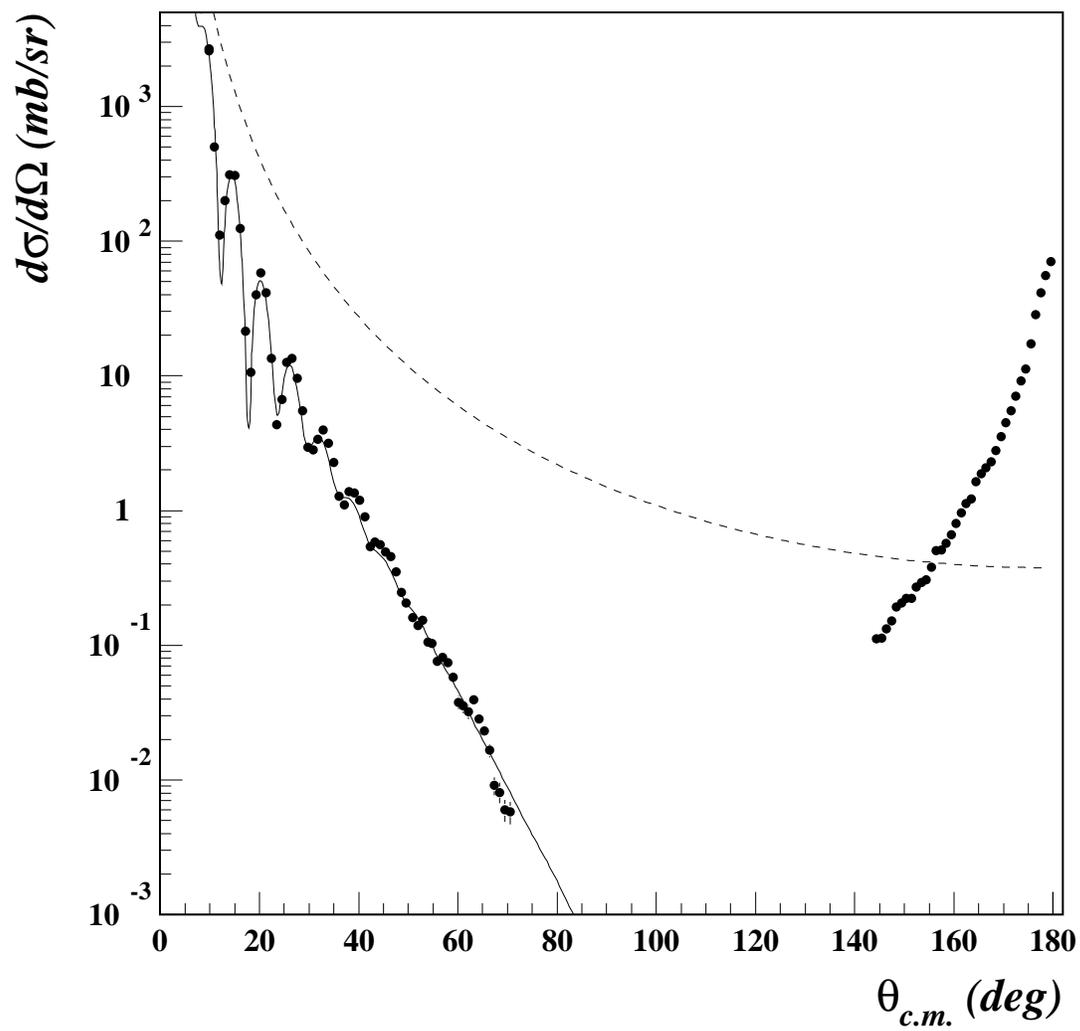
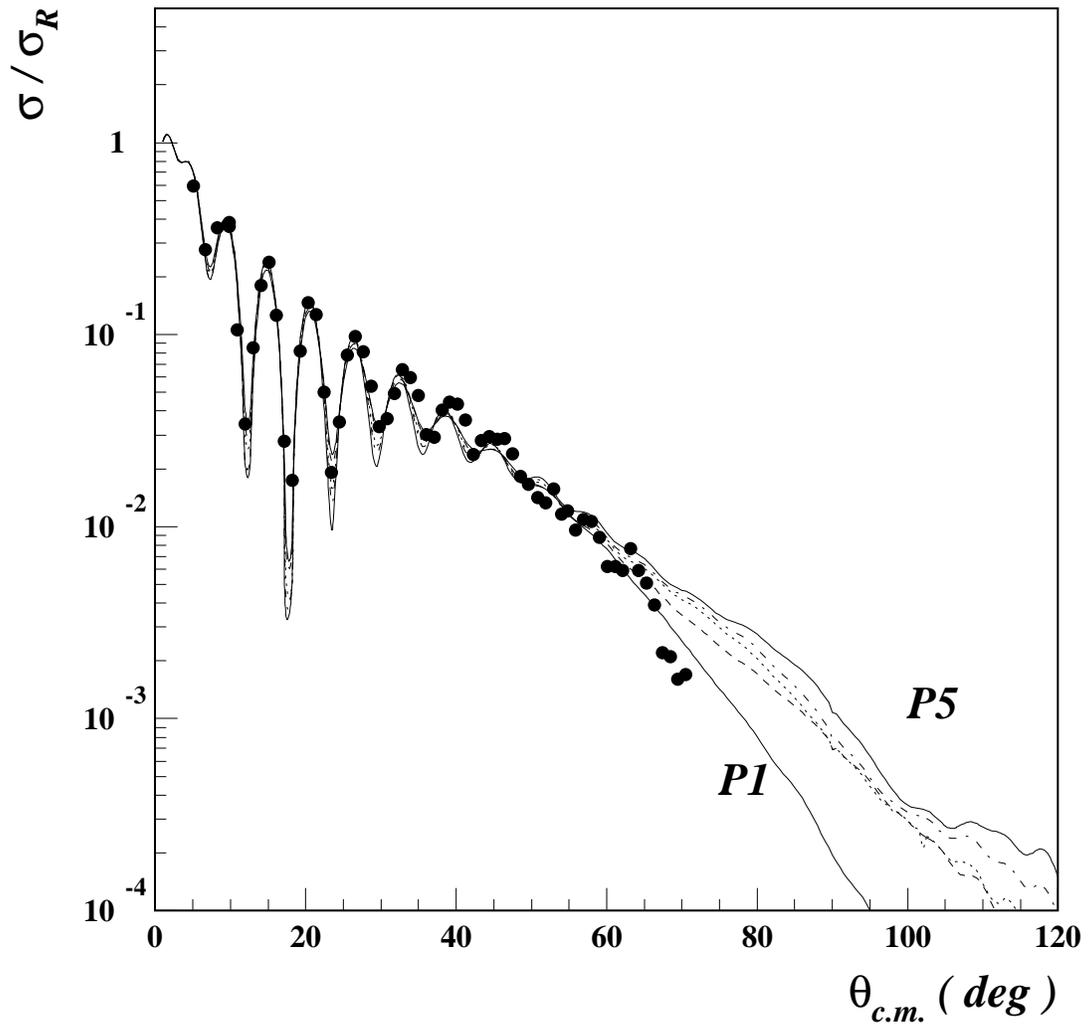
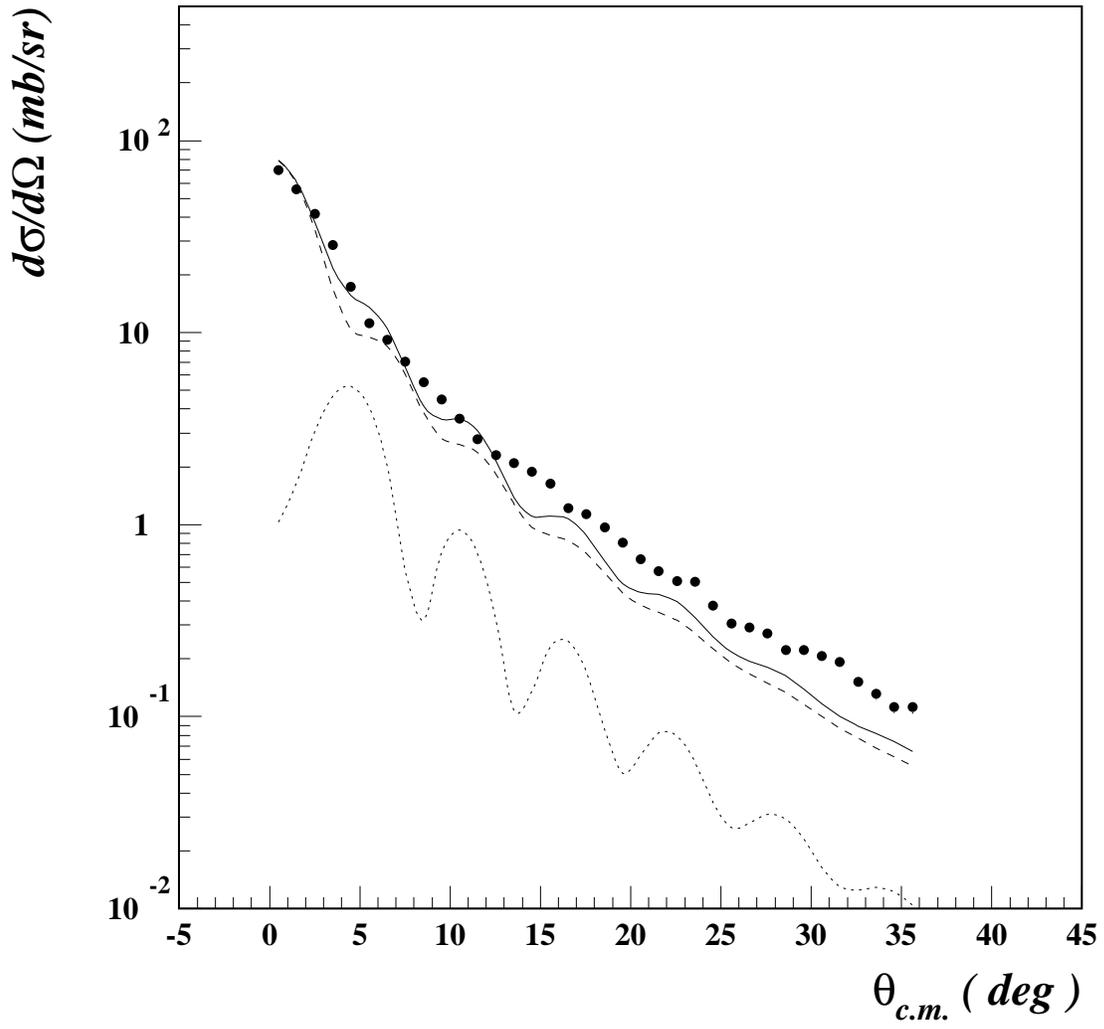
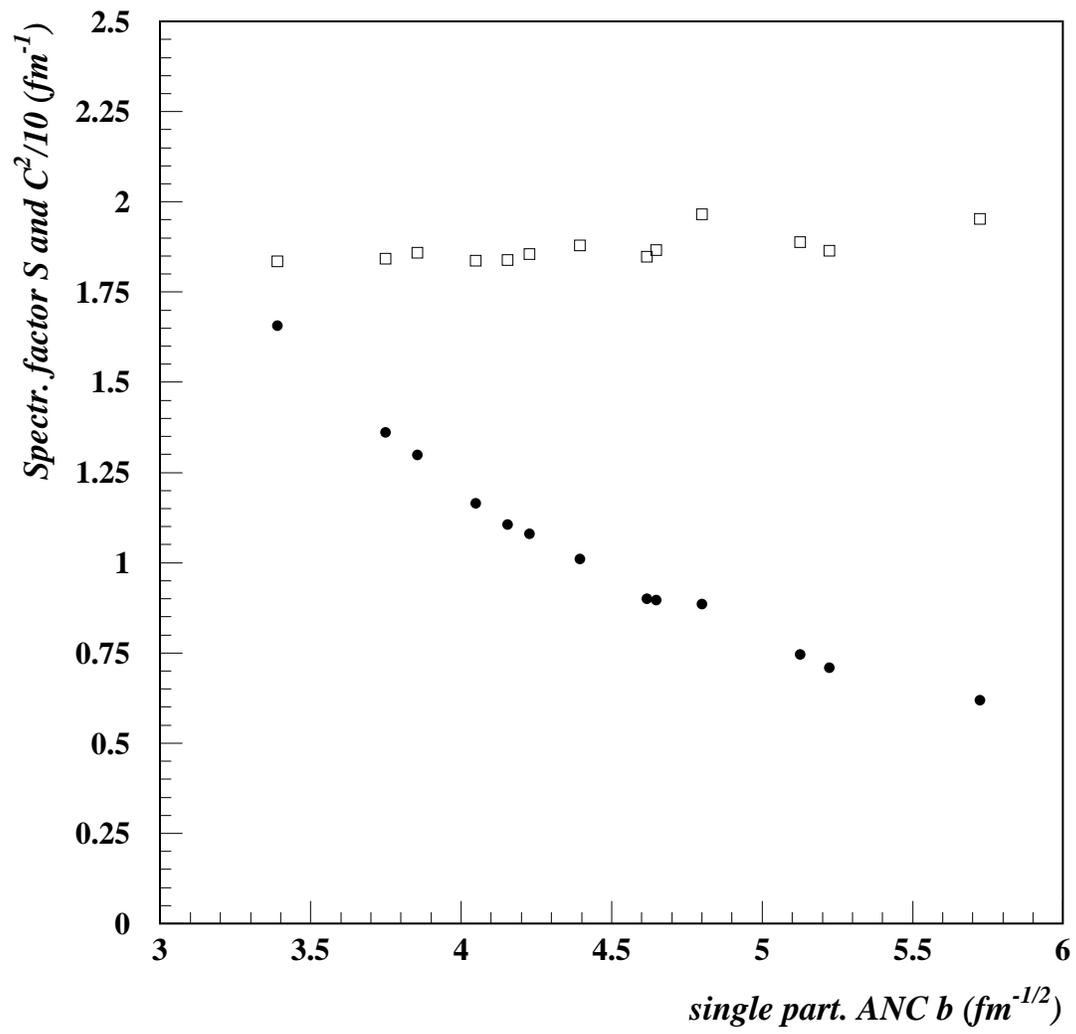
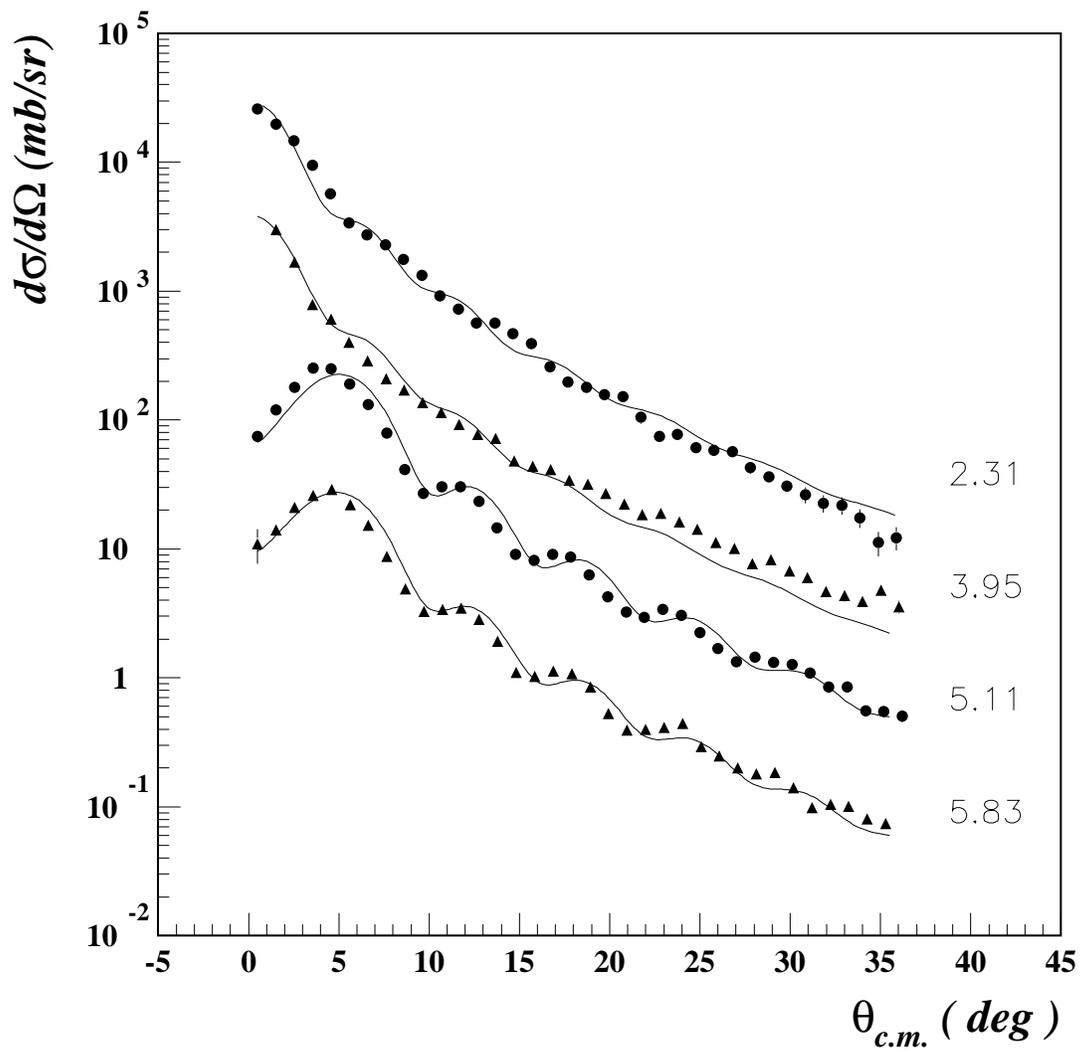
\end{document}